\documentclass[11pt, a4paper]{article}
\usepackage[T1]{fontenc}
\usepackage[utf8]{inputenc}
\usepackage{authblk}
\usepackage{amsmath,amsfonts,amssymb,amscd,amsthm,xspace}
\usepackage{rotating}
\usepackage{multirow}
\usepackage{subcaption}
\usepackage{graphicx}
\usepackage{hyperref}
\usepackage{vmargin}
\setmarginsrb  { 0.8in}  % left margin
               { 0.6in}  % top margin
               { 0.8in}  % right margin
               { 0.8in}  % bottom margin
               {  20pt}  % head height
               {0.25in}  % head sep
               {   9pt}  % foot height
               { 0.3in}  % foot sep

\title{Performance Analysis of Effective Symbolic Methods for Solving Band Matrix SLAEs}
\author[1]{Milena Veneva \thanks{milena.p.veneva@gmail.com}}
\author[1]{Alexander Ayriyan \thanks{ayriyan@jinr.ru}}
\affil[1]{Joint Institute for Nuclear Research, Laboratory of Information Technologies, Joliot-Curie 6, 141980 Dubna, Moscow region, Russia}

\date{}
\begin{document}
\maketitle
\abstract{%
This paper presents an experimental performance study of implementations of
three symbolic algorithms for solving band matrix systems of linear algebraic
equations with heptadiagonal, pentadiagonal, and tridiagonal coefficient
matrices. The only assumption on the coefficient matrix in order for the
algorithms to be stable is nonsingularity. These algorithms are implemented using the
\texttt{GiNaC} library of \texttt{C++} and the \texttt{SymPy} library of
\texttt{Python}, considering five different data storing classes. Performance
analysis of the implementations is done using the high-performance
computing~(HPC) platforms ``HybriLIT'' and ``Avitohol''. The experimental setup
and the results from the conducted computations on the individual computer
systems are presented and discussed. An analysis of the three algorithms is
performed.
}
\maketitle
%
%%%%%%%%%%%%%%%%%%%%%%%%%%%%%%%%%%%%%%%%%%%%%%%%%%%%%%%%%%%%%%%%%%%%%%%%%%%%%%%

\section{Introduction}
\label{intro}

Systems of linear algebraic equations~(SLAEs) with heptadiagonal~(HD),
pentadiagonal~(PD) and tridiagonal~(TD) coefficient matrices may arise after
many different scientific and engineering problems, as well as problems of the
computational linear algebra where finding the solution of a SLAE is
considered to be one of the most important problems. On the other hand, special
matrix’s characteristics like diagonal dominance, positive definiteness, etc.\,
are not always feasible. The latter two points explain why there is a need of
methods for solving of SLAEs which take into account the band structure of the
matrices and do not have any other special requirements to them. One
possible approach to this problem is the symbolic algorithms. An overview of
some of the symbolic algorithms which exist in the literature is done by us
in~\cite{ref_name_veneva_3}. What is common for all of them, is that they are
implemented using Computer Algebra Systems (CASs) such as \texttt{Maple},
\texttt{Mathematica}, and \texttt{Matlab}.

Three direct symbolic algorithms for solving a SLAE based on LU decomposition
are considered: for HD matrices (see \cite{ref_name_karawia} and
\cite{ref_name_veneva_3}) -- \textbf{SHDM}, for PD \cite{ref_name_askar} --
\textbf{SPDM}, and TD matrices \cite{ref_name_mik} -- \textbf{STDM}. The only
assumption on the coefficient matrix is nonsingularity.

The choice of algorithms for solving problems of the computational linear
algebra is crucial for the programs' effectiveness, especially when these
problems are with a big dimension. In that case they require the use of
supercomputers and computer clusters for the solution to be obtained in a
reasonable amount of time. Another important choice that has to be made and
that influences the programs' performance is what programming language to be
used for the algorithms' implementations. Here, we are going to focus on two of
the most popular programming languages for scientific computations, namely
\texttt{C++} and \texttt{Python}. The aim of this paper, which is a logical
continuation of \cite{ref_name_veneva_1} and \cite{ref_name_veneva_2}, is to
investigate the performance characteristics of the considered serial methods
with their two implementations being executed on modern computer clusters.

%%%%%%%%%%%%%%%%%%%%%%%%%%%%%%%%%%%%%%%%%%%%%%%%%%%%%%%%%%%%%%%%%%%%%%%%%%%%%%

\section{Computational Experiments}
\label{sec-3}

Computations were held on the basis of the heterogeneous computational platform
``HybriLIT'' (1142~TFlops/s for single precision and 550~TFlops/s for double
precision)~\cite{ref_name_hybrilit} at
the Laboratory of Information Technologies of the Joint Institute for Nuclear
Research in the town of science Dubna, Russia, and on the cluster computer system
``Avitohol'' (412.3~TFlops/s for double precision)~\cite{ref_name_avitohol} at
the Advanced Computing and Data Centre of the
Institute of Information and Communication Technologies of the Bulgarian Academy
of Sciences in Sofia, Bulgaria. The latter has been ranked among the TOP500 list
(\url{https://www.top500.org}) twice – being 332nd in June 2015 and 388th in
November 2015.

\subsection{Experimental Setup}

Table~\ref{tab:01} sums up some basic information about hardware on the two
computer systems, including models of processors, processors' base frequency,
and amount of cache memory
(SmartCache). For even more information, visit:
\url{https://ark.intel.com/compare/75281,75269}. Table~\ref{tab:02} summarizes
the basic information about the compilers and
libraries used on the two computer systems. The three algorithms are
implemented using the \texttt{GiNaC} library~\cite{ref_name_ginac} of
\texttt{C++}~\cite{ref_name_cpp_11} (the projects are built with the help of
\texttt{CMake}~\cite{ref_name_cmake}), and \texttt{SymPy}~\cite{ref_name_sympy}
library of \texttt{Python}~\cite{ref_name_python} (using Anaconda distribution~
\cite{ref_name_conda}). The reason why optimization -O0 was used is that the
\texttt{GiNaC} library is already optimized and any further attempts to
optimize the code give worse results. Although we use Py~2.7, one should note
that the implementations are fully compatible with Py~3.6 as well. It is worth
mentioning that optimization attempts like \texttt{autowrap} and
\texttt{numba} do not give better execution times. The former because of
overhead, the second one because it cannot optimize further than what is already
done.
%There is hope that if the Google Summer of Code's project \texttt{sympyx}
%(\texttt{SymPy} core written in \texttt{Cython}, for more information, visit
%\url{https://github.com/certik/sympyx/})
%is completed someday, its usage might give better results.

\renewcommand{\arraystretch}{1.2}
\begin{table}[!htb]
\centering
\caption{Intel processors used for numerical experiments.}
\label{tab:01}
\begin{tabular}{cccc}
\hline\hline
Computer system & Processor & FREQ [GHz] & Cache [MB] \\\hline\hline
``HybriLIT'' & Intel Xeon E5-2695v2 & 2.40 & 30 \\\hline
``Avitohol'' & Intel Xeon E5-2650v2 & 2.60 & 20 \\\hline\hline       			
\end{tabular}
\end{table}
\renewcommand{\arraystretch}{1.2}
\begin{table}[!htb]
\centering
\caption{Information about the used software on the two computer systems.}
\label{tab:02}
\begin{tabular}{c|r|c|c}
\hline\hline
\multicolumn{2}{r|}{Computer}& ``HybriLIT''          & ``Avitohol''   \\
\multicolumn{2}{r|}{system}  &                       &            \\\hline
\multicolumn{2}{r|}{OS}       & Scientific Linux 7.4  & Red Hat Linux \\\hline\hline
\multirow{4}{*}{\rotatebox[origin=c]{90}{C++}}
            & Compilers     & GCC (4.9.3) & GCC (6.2.0) \\\cline{2-4}
            & \multirow{2}{*}{Libraries}       & GiNaC (1.7.3) & GiNaC (1.7.2) \\\cline{3-4}
            &                    			   & \multicolumn{2}{c}{CLN (1.3.4)}\\\cline{2-4}
            & Optimization & \multicolumn{2}{c}{-O0}         \\\hline\hline
\multirow{2}{*}{\rotatebox[origin=c]{90}{\small Python}}
            & Version & \multicolumn{2}{c}{Anaconda (5.0.1): Py2.7}\\\cline{2-4}
            & Library & \multicolumn{2}{c}{SymPy (1.1.1)} \\\hline\hline
\end{tabular}
\end{table}

\subsection{Experimental Results and Analysis of the Algorithms}

During our experiments wall-clock times were collected and the average time
from multiple runs is reported. For that purpose, we use 
\texttt{std::chrono::high\_resolution\_clock::now()} for
\texttt{C++} (requires at least standard \texttt{c++11}), and
\texttt{timeit.default\_timer()} for \texttt{Python}. Both of them provide the
best clock rate available on the platform. Five different classes for data
storing are tested -- \texttt{lst} and \texttt{matrix} of \texttt{GiNaC};
\texttt{Matrix} (variable-size), \texttt{Matrix} (fixed-size), and
\texttt{MutableDenseNDimArray} of
\texttt{SymPy}. The first and the third are intended for variable-size storing,
while the others -- for fixed-size one (the SymPy's Matrix could be both).
Further we are going to denote the implementations of the three algorithms,
using these 5 data storing classes as Impl.\,i, $i=\overline{1,\ldots 5}$.
The notation is as follows: \textbf{SXDM} stands for symbolic X method,
X $=$ {HD, PD, TD}. The achieved computational times from solving a
SLAE are summarized in Tables~\ref{tab:03}--\ref{tab:07}.

\begin{table}[!htb]
\centering
\caption{Results from solving a SLAE on the two clusters applying the first implementation.}
\label{tab:03}
\begin{tabular}{crrr||rrr}
\hline\hline
& \multicolumn{6}{c}{GiNaC (``lst'' -- var-size) := Impl.\,1}  \\\hline\hline
& \multicolumn{6}{c}{Wall-clock time [s]}  \\\hline\hline
& \multicolumn{3}{c||}{``HybriLIT''} & \multicolumn{3}{c}{``Avitohol''} \\\hline\hline
$N$ & SHDM & SPDM & \multicolumn{1}{r||}{STDM} & SHDM & SPDM & STDM	 \\\hline
$10^{3}$ & 0.191933 & 0.120892 & 0.054486 & 0.211552 & 0.108980 & 0.051806 \\
$10^{4}$ & 29.346658 & 14.384663 & 6.4003244 & 30.525208 & 15.238341 & 5.280648 \\
$10^{5}$ & 5095.508994 & 2389.583291 & 805.592624 & 4026.259486 & 2009.600485 & 711.940280\\
\hline\hline
\end{tabular}
\end{table}
\vspace{-0em}
\begin{table}[!htb]
\centering
\caption{Results from solving a SLAE on the two clusters applying the second implementation.}
\label{tab:04}
\begin{tabular}{crrr||rrr}
\hline\hline
& \multicolumn{6}{c}{GiNaC (``matrix'' -- fixed-size) := Impl.\,2}  \\\hline\hline
& \multicolumn{6}{c}{Wall-clock time [s]}  \\\hline\hline
& \multicolumn{3}{c||}{``HybriLIT''} & \multicolumn{3}{c}{``Avitohol''} \\\hline\hline
$N$ & SHDM & SPDM & \multicolumn{1}{r||}{STDM} & SHDM & SPDM & STDM	 \\\hline
$10^{3}$ & 0.025110 & 0.016281 & 0.008808 & 0.030020 & 0.017507 & 0.010651 \\
$10^{4}$ & 0.254571 & 0.153711 & 0.086114 & 0.296938 & 0.173316 & 0.091560\\
$10^{5}$ & 2.567423 & 1.553468 & 0.858090 & 2.822281 & 1.694611 & 0.888428\\
\hline\hline
\end{tabular}
\end{table}
\vspace{-0em}
\begin{table}[!htb]
\centering
\caption{Results from solving a SLAE on the two clusters applying the third implementation.}
\label{tab:05}
\begin{tabular}{crrr||rrr}
\hline\hline
& \multicolumn{6}{c}{SymPy (``Matrix'' -- var-size) := Impl.\,3}  \\\hline\hline
& \multicolumn{6}{c}{Wall-clock time [s]}  \\\hline\hline
& \multicolumn{3}{c||}{``HybriLIT''} & \multicolumn{3}{c}{``Avitohol''} \\\hline\hline
$N$ & SHDM & SPDM & \multicolumn{1}{r||}{STDM} & SHDM & SPDM & STDM	 \\\hline
$10^{2}$ & 0.779426 & 0.541518 & 0.288752 & 1.198145 & 0.854763 & 0.450166 \\
$10^{3}$ & 61.955375 & 47.205663 & 23.203961 & 102.278640 & 75.408729 & 36.983316 \\
$10^{4}$ & 6159.227881 & 4587.017868 & 2294.136577 & 9967.937595 & 7517.447892 & 3751.046286 \\
\hline\hline
\end{tabular}
\end{table}
\vspace{-0em}
\begin{table}[!htb]
\centering
\caption{Results from solving a SLAE on the two clusters applying the fourth implementation.}
\label{tab:06}
\begin{tabular}{crrr||rrr}
\hline\hline
& \multicolumn{6}{c}{SymPy (``Matrix'' -- fixed-size) := Impl.\,4}  \\\hline\hline
& \multicolumn{6}{c}{Wall-clock time [s]}  \\\hline\hline
& \multicolumn{3}{c||}{``HybriLIT''} & \multicolumn{3}{c}{``Avitohol''} \\\hline\hline
$N$ & SHDM & SPDM & \multicolumn{1}{r||}{STDM} & SHDM & SPDM & STDM	 \\\hline
$10^{3}$ & 0.875309 & 0.726237 & 0.417078 & 1.272449 & 0.788081 & 0.429694 \\
$10^{4}$ & 8.420457 & 5.780376 & 2.909977 & 12.807111 & 7.738780 & 4.163307 \\
$10^{5}$ & 84.977702 & 60.360366 & 29.446712 & 128.940536 & 77.026990 & 41.930210 \\
\hline\hline
\end{tabular}
\end{table}
\vspace{-0em}
\begin{table}[!htb]
\centering
\caption{Results from solving a SLAE on the two clusters applying the fifth implementation.}
\label{tab:07}
\begin{tabular}{crrr||rrr}
\hline\hline
& \multicolumn{6}{c}{SymPy (``MutableDenseNDimArray'' -- fixed-size) := Impl.\,5}  \\\hline\hline
& \multicolumn{6}{c}{Wall-clock time [s]}  \\\hline\hline
& \multicolumn{3}{c||}{``HybriLIT''} & \multicolumn{3}{c}{``Avitohol''} \\\hline\hline
$N$ & SHDM & SPDM & \multicolumn{1}{r||}{STDM} & SHDM & SPDM & STDM	 \\\hline
$10^{3}$ & 0.730619 & 0.545021 & 0.308278 & 1.104916 & 0.694794 & 0.387745 \\
$10^{4}$ & 7.472471 & 4.941960 & 3.115306 & 11.105768 & 6.726764 & 3.759721 \\
$10^{5}$ & 73.927031 & 51.304943 & 30.917188 & 111.968824 & 67.404472 & 37.466753 \\
\hline\hline
\end{tabular}
\end{table}

\renewcommand{\arraystretch}{1.2}
\begin{table}[!htb]
\centering
\caption{Estimation of the order $\alpha(10^4,10^5)$.}
\label{tab:11}
\begin{tabular}{rccccc||ccccc}
\hline\hline
\multicolumn{11}{c}{$\alpha(10^4,10^5)$}  \\\hline\hline
     & \multicolumn{5}{c||}{``HybriLIT''} & \multicolumn{5}{c}{``Avitohol''} \\\hline\hline
Impl.& \#1 & \#2 & \#3 & \#4 & \#5 & \#1 & \#2 & \#3 & \#4 & \#5\\\hline\hline
SHDM & 2.24 & 1.00 & 2.00 & 1.00 & 1.00 & 2.12 & 0.98 & 1.99 & 1.00 & 1.00\\\hline
SPDM & 2.22 & 1.00 & 1.99 & 1.02 & 1.02 & 2.12 & 0.99 & 2.00 & 1.00 & 1.00  \\\hline
STDM & 2.10 & 1.00 & 2.00 & 1.01 & 1.00 & 2.13 & 0.99 & 2.01 & 1.00 & 1.00  \\\hline\hline      			
\end{tabular}
\end{table}

Using the following formula ($k$ -- unknown coefficient of proportionality,
$t_{i}$ -- time, $N_{i}$ -- the matrix's number of rows, $i = 1, 2$):
\begin{equation*}
t\approx kN^{\alpha}\quad\Rightarrow\quad\frac{t_{2}}{t_{1}} = \left(\frac{N_{2}}{N_{1}}\right)^{\alpha}\quad\Leftrightarrow\quad\alpha(N_{1}, N_{2}) = %\frac{\log\left(\frac{t_{2}}{t_{1}}\right)}{\log\left(\frac{N_{2}}{N_{1}}\right)} =
\dfrac{\log(t_{2}) - \log(t_{1})}{\log(N_{2}) - \log(N_{1})},
\end{equation*}
the order of growth of execution time $\alpha$ for all the five implementations
was estimated (see Table~\ref{tab:11}).

\begin{figure}[!htb]
\centering
\begin{subfigure}{0.5\textwidth}
  \centering
  \includegraphics[width=\linewidth]{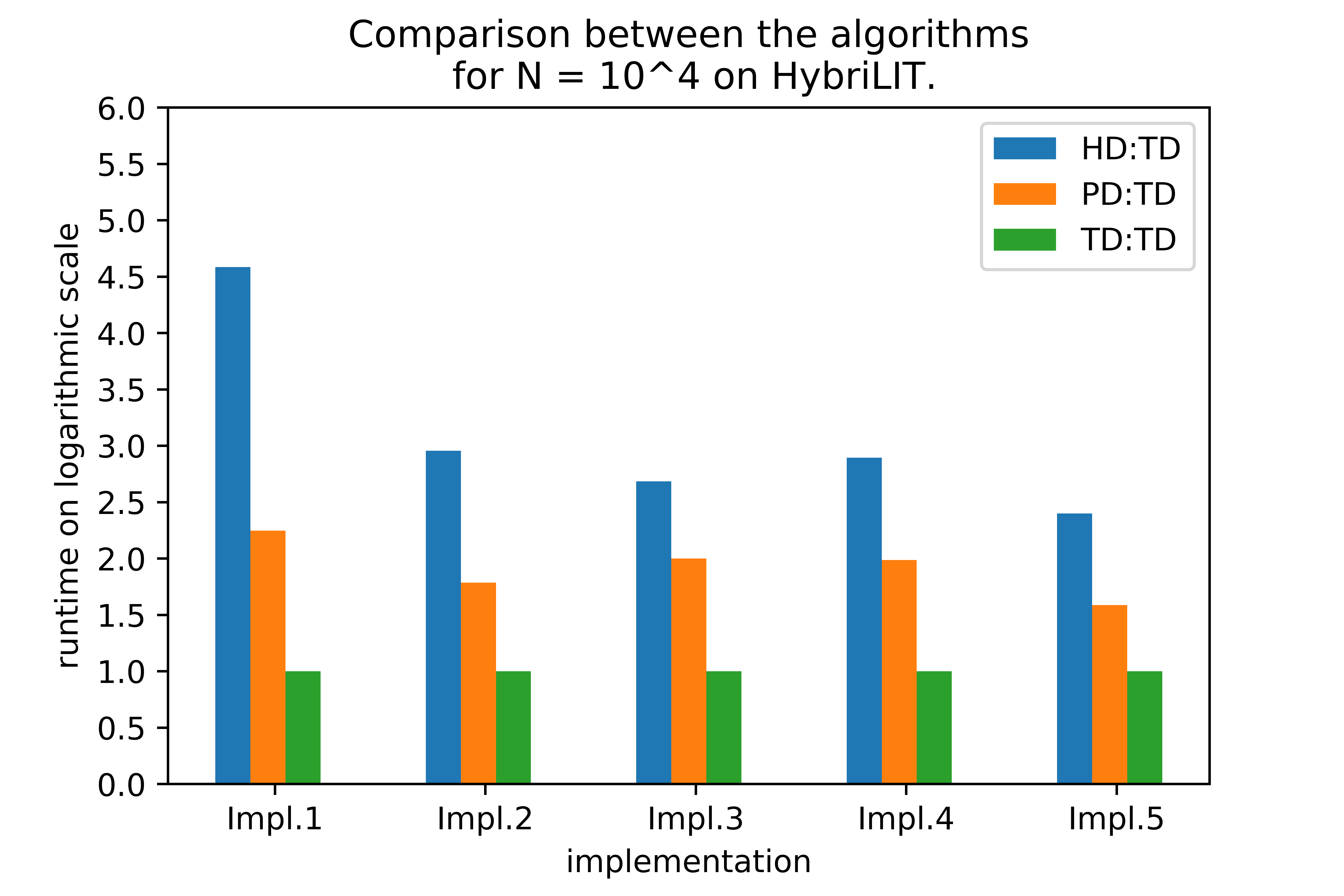}
  %\caption{.}
\end{subfigure}%
\begin{subfigure}{0.5\textwidth}
  \centering
  \includegraphics[width=\linewidth]{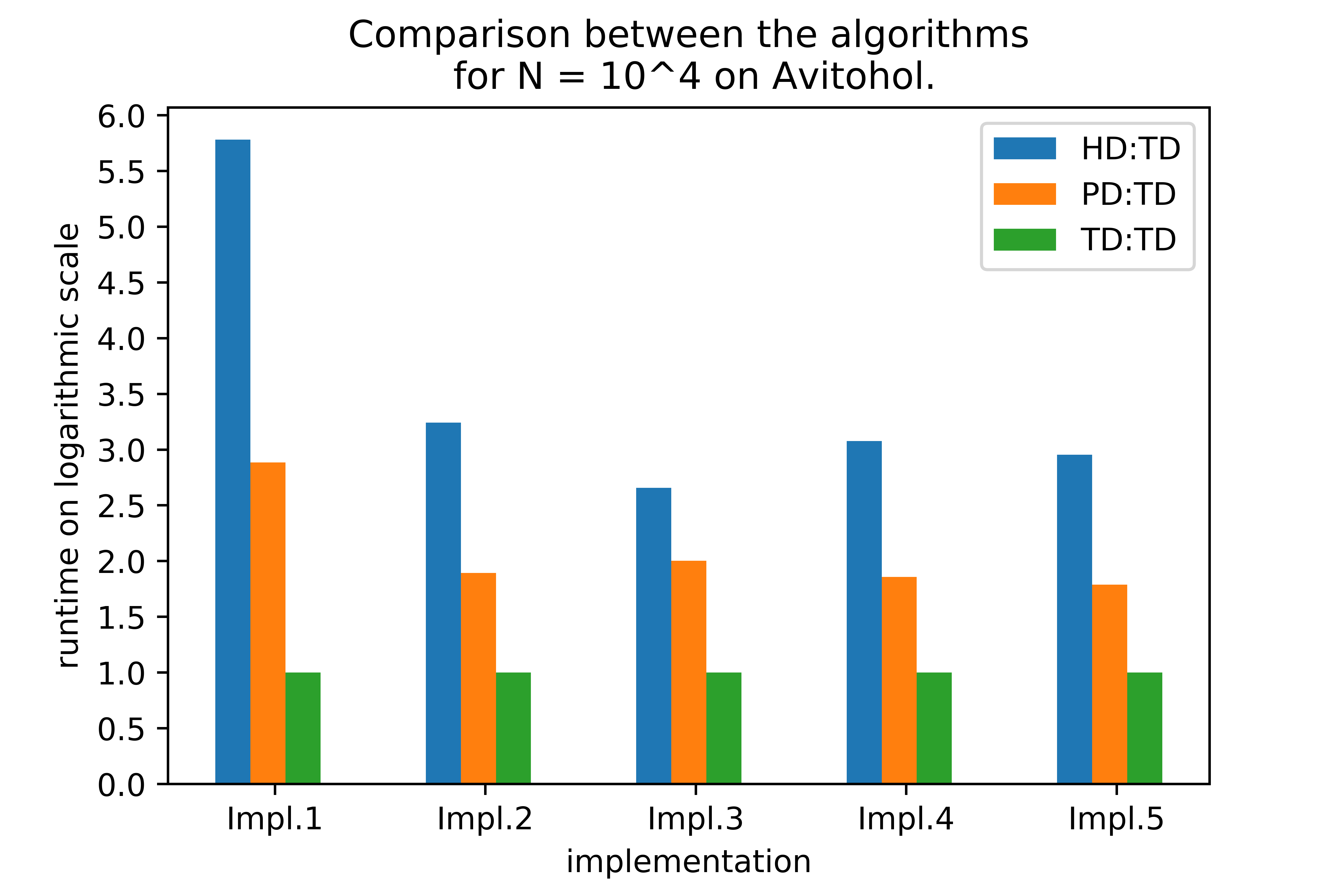}
  %\caption{.}
\end{subfigure}
\caption{Comparison between the algorithms for $N = 10^4$.}
\label{fig:4}
\end{figure}

\textbf{Remark:} The number of needed operations for Gaussian elimination so
as a HD SLAE to be transformed into a PD one is $35N-122$, while the number of
needed operations so as a PD SLAE to be transformed into a TD one is $23N-52$,
where $N$ is the matrix's number of rows. This observation is relevant to
Figure~\ref{fig:4} which depicts the factors of time growth HD:TD, PD:TD, TD:TD
for each of the implementations.

%%%%%%%%%%%%%%%%%%%%%%%%%%%%%%%%%%%%%%%%%%%%%%%%%%%%%%%%%%%%%%%%%%%%%%%%%%%%%%%

\section{Discussion and Conclusions}
\label{sec-4}

\begin{figure}[!htb]
\centering
\begin{subfigure}{0.5\textwidth}
  \centering
  \includegraphics[width=\linewidth]{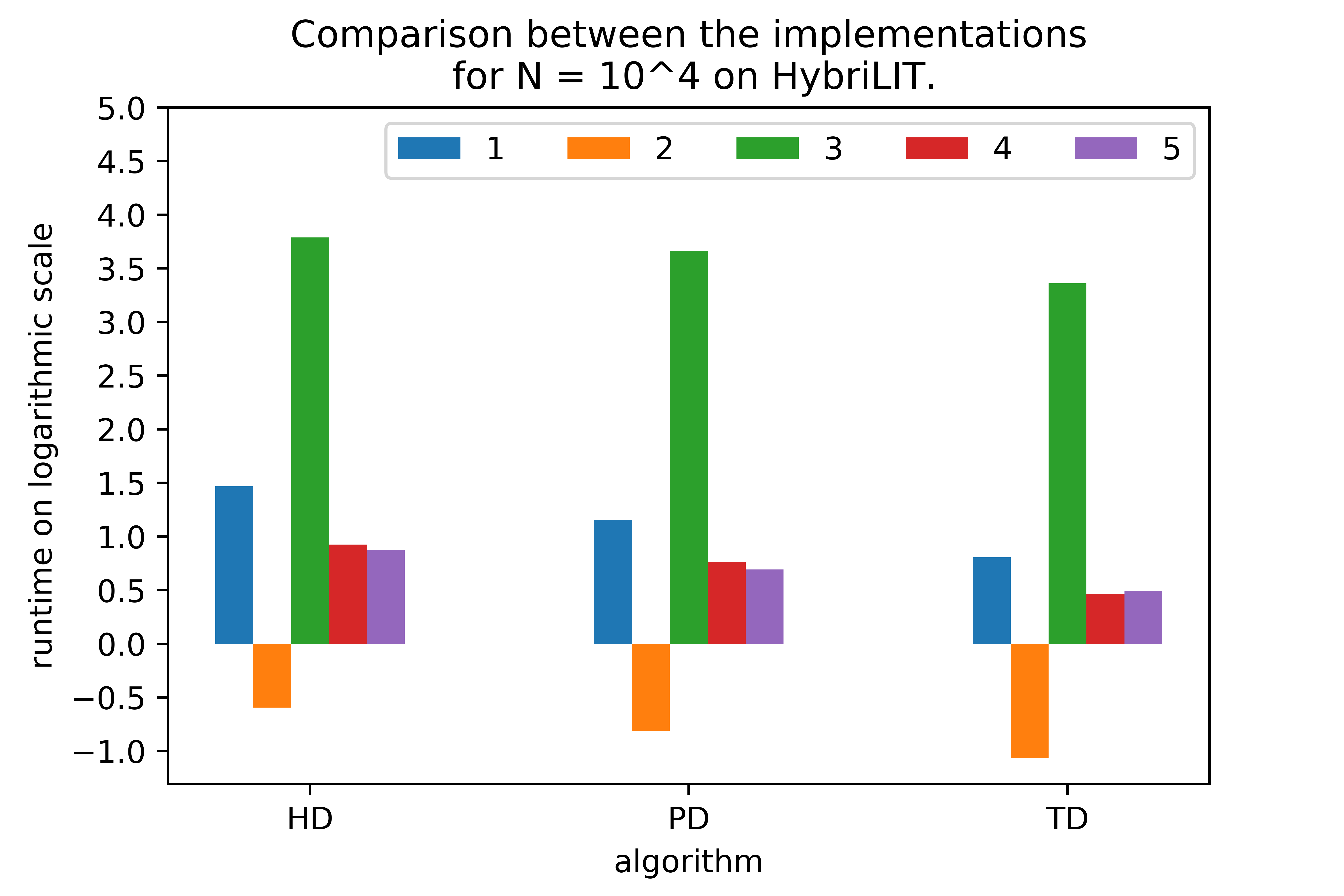}
\end{subfigure}%
\begin{subfigure}{0.5\textwidth}
  \centering
  \includegraphics[width=\linewidth]{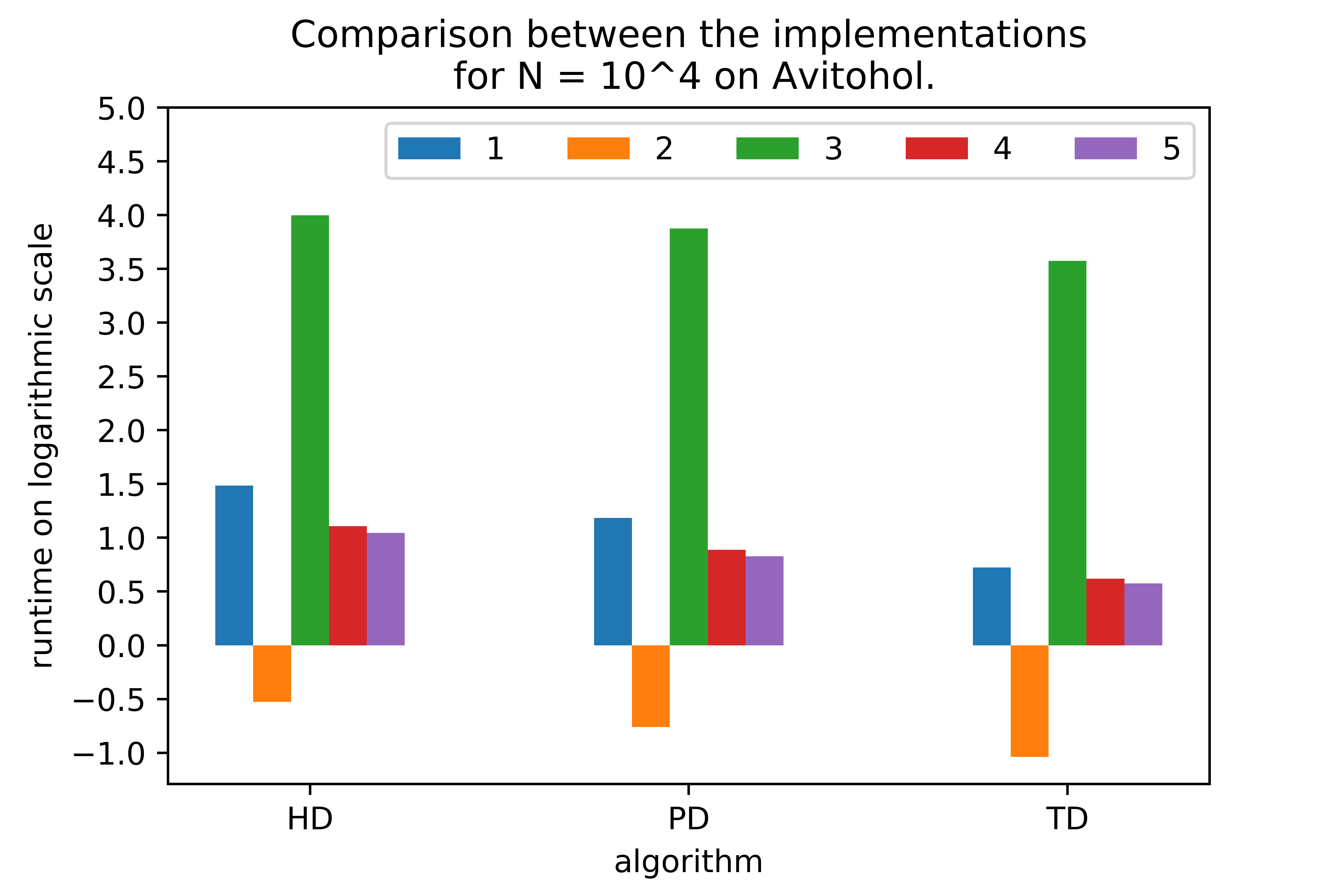}
\end{subfigure}
\caption{Comparison between the implementations for $N = 10^4$.}
\label{fig:3}
\end{figure}

\begin{figure}[!htb]
\centering
\begin{subfigure}{0.5\textwidth}
  \centering
  \includegraphics[width=\linewidth]{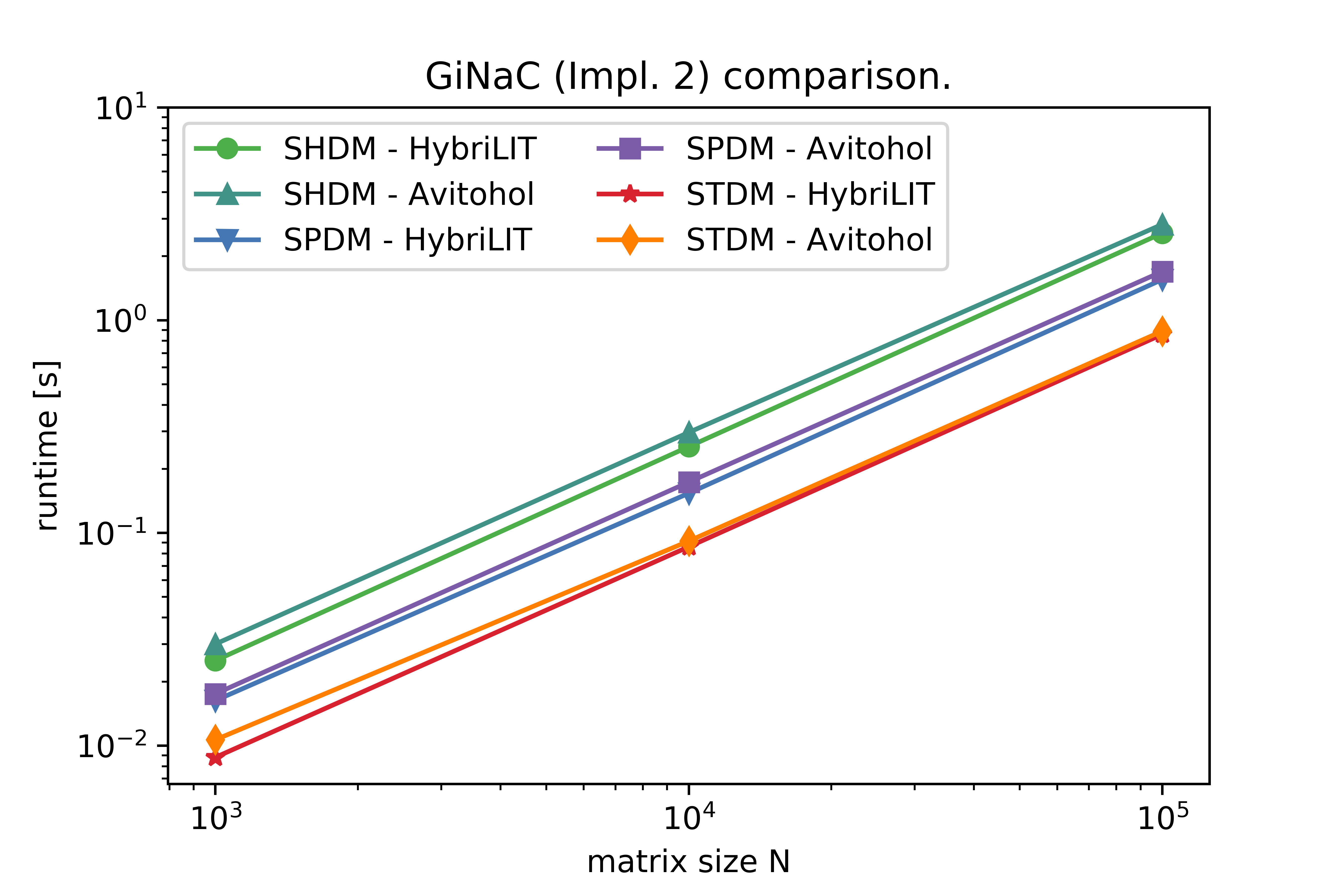}
  %\caption{.}
\end{subfigure}%
\begin{subfigure}{0.5\textwidth}
  \centering
  \includegraphics[width=\linewidth]{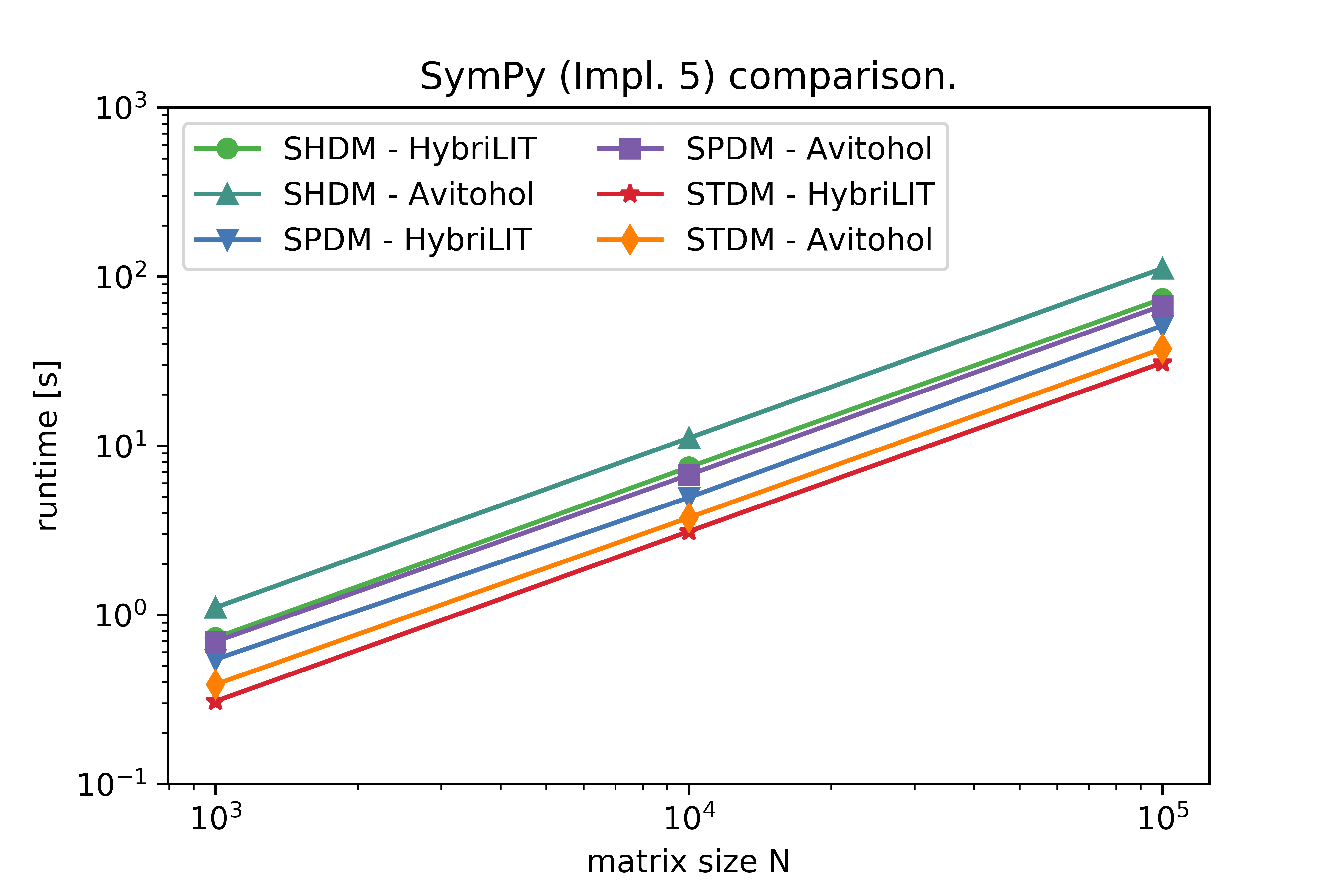}
  %\caption{.}
\end{subfigure}
\caption{Comparison of the execution time of the best \texttt{GiNaC} and
\texttt{SymPy} implementations on the two clusters.}
\label{fig:1}
\end{figure}

\begin{figure}[!htb]
\centering
\begin{subfigure}{0.5\textwidth}
  \centering
  \includegraphics[width=\linewidth]{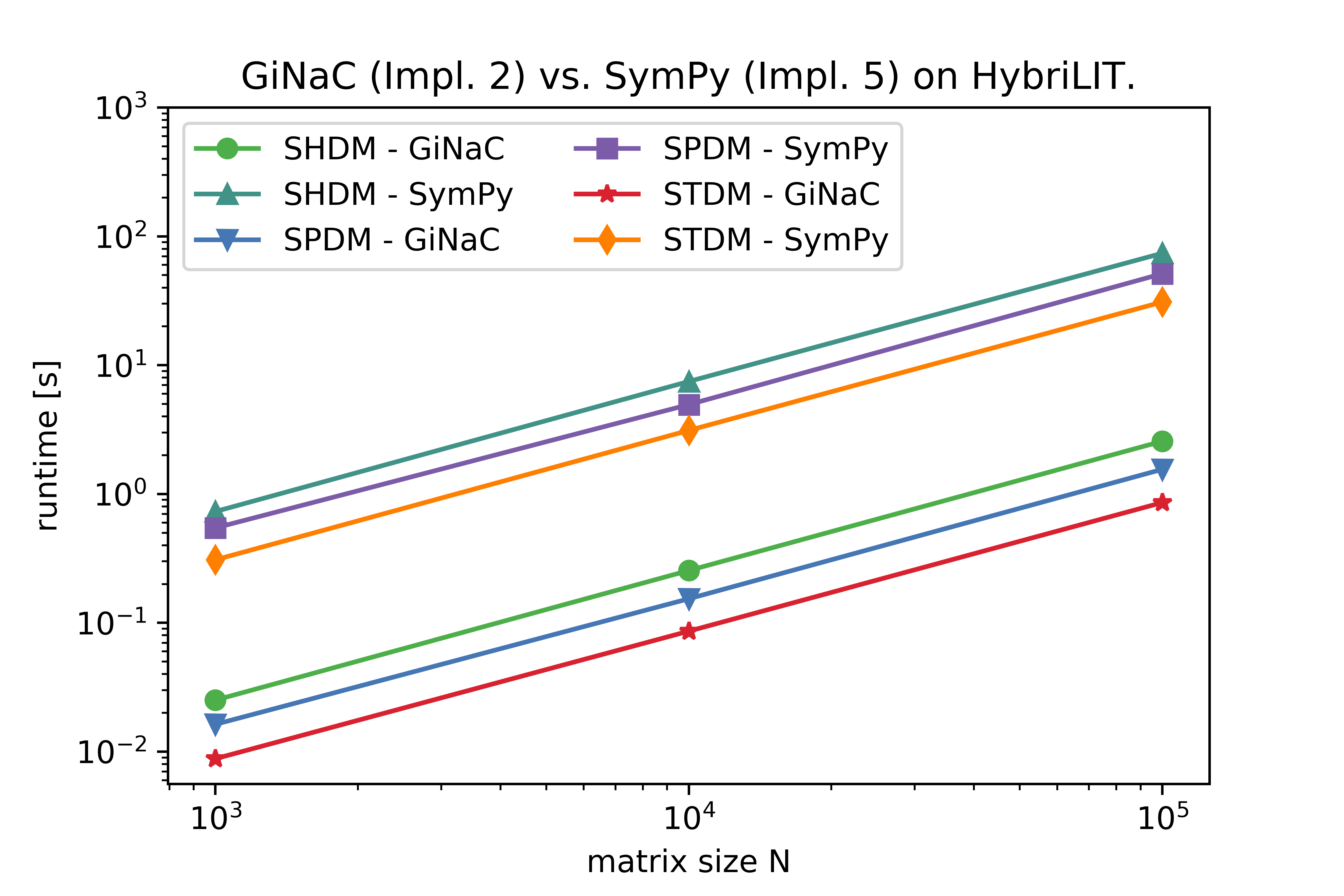}
  %\caption{.}
\end{subfigure}%
\begin{subfigure}{0.5\textwidth}
  \centering
  \includegraphics[width=\linewidth]{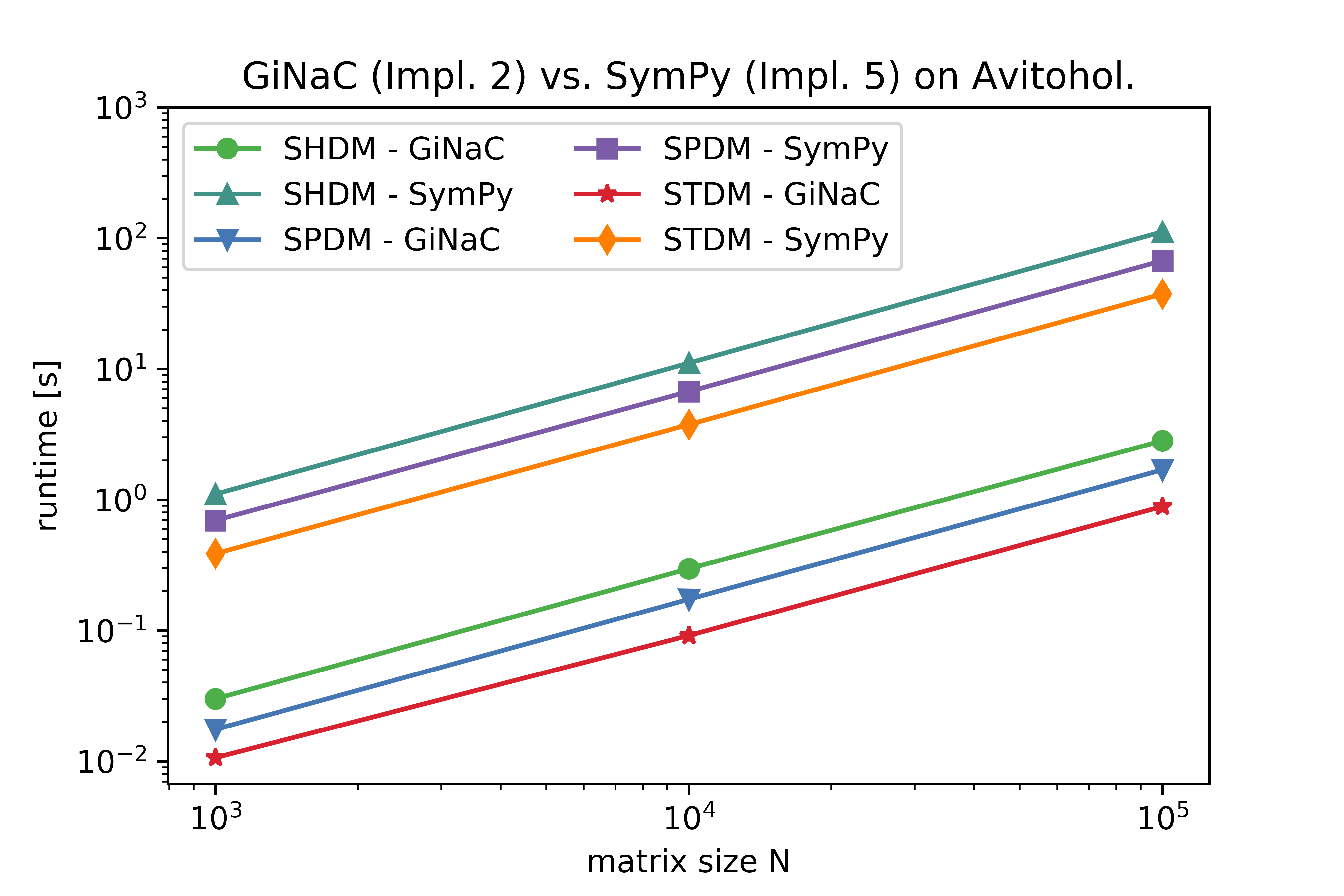}
  %\caption{.}
\end{subfigure}
\caption{Comparison of the execution time of the best \texttt{GiNaC} and
\texttt{SymPy} implementations on the two clusters.}
\label{fig:2}
\end{figure}

Direct comparison between the five implementations shows that the best
\texttt{GiNaC} one is Impl.\,2, while the best \texttt{SymPy} one is Impl.\,5
(see Figure~\ref{fig:3}). Expectedly, the \texttt{GiNaC} implementations of the
three algorithms yield a much better computation time in comparison with the
respected \texttt{SymPy} implementations with the difference being between
one and two orders of magnitude. It must be mentioned that the matrix class in
\texttt{SymPy} is a subclass of the ndarray. This means that every call on a
matrix object requires a few extra \texttt{Python} calls. This usually leads to
a little slower performance although the difference is negligible.
Figures~\ref{fig:1} and \ref{fig:2} depict a comparison of the execution time of
the best \texttt{GiNaC} and \texttt{SymPy} implementations on the two clusters.
As one can see, ``HybriLIT'' behaves better than ``Avitohol'' with the
difference being bigger when the \texttt{SymPy} is of interest. It is obvious
that the implementations which rely on variable-size storing classes (that are
Impl.\,2 and Impl.3) were found to be much slower than the ones which use
fixed-size storing classes, but while this was expected, it does not belittle
their importance since there is a class of problems where the size of the
matrix is unknown at runtime.
\vspace{-0em}
\vspace{-0em}

As it can be seen in Table~\ref{tab:10}, three of the implementations have a
linear time growth, while the other two have a quadratic trend. Theoretically,
all the algorithms introduced here have a linear complexity. This means that
even if theoretically the execution time has to grow as $O(N)$, the two
implementations which rely on variable-size storing classes show time growth as
$O(N^2)$, with Impl.\,1 even having $\alpha > 2$. Hence, the variable-size of
the storing class changes the order of complexity with one order of magnitude.
\renewcommand{\arraystretch}{1.2}
\begin{table}[!htb]
\centering
\caption{Order of time growth.}
\label{tab:10}
\begin{tabular}{cc}
\hline\hline
Order of growth & Implementation \\\hline\hline
$O(N)$     & 2, 4, 5 \\\hline
$O(N^{2})$ & 1, 3 \\\hline\hline      			
\end{tabular}
\end{table}

The choice of a programming language depends on a lot of factors. However, in the
context of symbolic computations for solving a SLAE with a band coefficient
matrix (length of band equal to 3, 5 or 7) among the options suggested in this work, we
can note the following: \texttt{Python} is easier to learn and easier and
faster to prototype (no need for memory management, build systems, compilers,
etc.), on the other hand, \texttt{C++} has better performance and occupies
less memory, but requires much more attention to bookkeeping and storage details.

%Which programming language to choose for the sake of solving a SLAE with a band
%coefficient matrix (length of band equal to 3, 5 or 7) among the suggested in
%this note depends on a lot of factors, the most important probably being the
%project.
%
%\texttt{Python} is easier to learn and easier and faster to
%prototype (no need for memory management, build systems, compilers, etc.). On
%the other hand, \texttt{C++} has better performance and occupies less memory,
%but requires much more attention to bookkeeping and storage details.
%
%All the figures in this paper have been generated using
%\texttt{Matplotlib}~(2.1.0) \cite{ref_name_matplotlib}.
%
%\vspace{-0em}
%\begin{figure}[!htb]
%  \sidecaption
%  \centering
%  \includegraphics[width=0.6\textwidth]{ginac_comp_new_lst.png}
%  \caption{Comparison of the execution time for Impl.\,1.}
%  \label{fig:5}
%\end{figure}
%%%%%%%%%%%%%%%%%%%%%%%%%%%%%%%%%%%%%%%%%%%%%%%%%%%%%%%%%%%%%%%%%%%%%%%%%%%%%%%
\section*{Acknowledgements}
%\begin{acknowledgement}
The authors want to express their gratitude to the Summer Student
Program at JINR, Dr.\,J\'{a}n Bu\v{s}a Jr. (JINR), Dr.\,Andrey
Lebedev (GSI/JINR), Assoc.\,Prof.\,Ivan Georgiev (IICT \& IMI,
BAS), the ``HybriLIT'' team at LIT, JINR, and the ``Avitohol'' team
at the Advanced Computing and Data Centre of IICT, BAS. Computer
time grants from LIT, JINR and the Advanced Computing and Data
Centre at IICT, BAS are kindly acknowledged. The work is partially
supported by the Russian Foundation for Basic Research under project
\#18-51-18005. All the figures in this paper have been generated using
\texttt{Matplotlib}~(2.1.0) \cite{ref_name_matplotlib}.
%\end{acknowledgement}

%%%%%%%%%%%%%%%%%%%%%%%%%%%%%%%%%%%%%%%%%%%%%%%%%%%%%%%%%%%%%%%%%%%%%%%%%%%%%%


\begin{thebibliography}{99.}
\bibitem{ref_name_veneva_3}
    Veneva, M., Ayriyan, A.: Symbolic Algorithm for Solving SLAEs with
    Heptadiagonal Coefficient Matrices. Mathematical Modelling and Geometry,
    \textbf{6}, 3, 22--29 (2018).
\bibitem{ref_name_karawia} Karawia, A.~A.: A New Algorithm for General
    Cyclic Heptadiagonal Linear Systems Using Sherman-Morrisor-Woodbury Formula.
    {ARS} Combinatoria, \textbf{108}, 431--443 (2013).
\bibitem{ref_name_askar}
    Askar, S.~S., Karawia, A.~A.: On Solving Pentadiagonal Linear Systems via
    Transformations. Mathematical Problems in Engineering. Hindawi Publishing
    Corporation. \textbf{2015}, 9 (2015), doi:~10.1155/2015/232456.
\bibitem{ref_name_mik}
    El-Mikkawy, M.: A Generalized Symbolic Thomas Algorithm. Applied
    Mathematics. \textbf{3}, 4, 342--345 (2012), doi:~10.4236/am.2012.34052.
\bibitem{ref_name_veneva_1}
    Veneva, M., Ayriyan, A.: Effective Methods for Solving Band SLAEs after
    Parabolic Nonlinear PDEs. AYSS-2017, European Physics Journal -- Web of
    Conferences~(EPJ-WoC). \textbf{177}, 07004 (2018).
\bibitem{ref_name_veneva_2}
    Veneva, M., Ayriyan, A.: Performance Analysis of Effective Methods for
    Solving Band Matrix SLAEs after Parabolic Nonlinear PDEs. Advanced
    Computing in Industrial Mathematics, Revised Selected Papers of the 12th
    Annual Meeting of the Bulgarian Section of SIAM, December 20--22, 2017,
    Sofia, Bulgaria, Studies in Computational Intelligence, \textbf{793},
    407--419 (2019), doi:~10.1007/978-3-319-97277-0\_33.
\bibitem{ref_name_hybrilit}
    Adam, Gh., Bashashin, M., Belyakov, D., Kirakosyan, M., Matveev, M., Podgainy, D.,
    Sapozhnikova, T., Streltsova, O., Torosyan, Sh., Vala, M., Valova, L.,
    Vorontsov, A., Zaikina, T., Zemlyanaya, E., Zuev, M.: {IT}-ecosystem
    of the {HybriLIT} Heterogeneous Platform for High-performance Computing
    and Training of {IT}-specialists. Selected Papers of the 8th International
    Conference ``Distributed Computing and Grid-technologies in Science and
    Education'' (GRID~2018), Dubna, Russia, September 10--14, 2018, \textbf{2267},
    638--644.
\bibitem{ref_name_avitohol}
    Supercomputer System Avitohol at IICT-BAS, \url{http://www.hpc.acad.bg/}.
\bibitem{ref_name_ginac}
    Bauer, C., Frink, A., Kreckel, R.: Introduction to the {GiNaC} Framework
    for Symbolic Computation within the C++ Programming Language. J. Symbolic
    Computation. \textbf{33}, 1--12 (2002), doi:~10.1006/jsco.2001.0494.
\bibitem{ref_name_cpp_11}
    ISO/IEC. (2012). ISO International Standard ISO/IEC for Programming Language
    {C++}. [Working draft], N3337. Geneva, Switzerland: International Organization
    for Standardization~(ISO). Retrieved from \url{https://isocpp.org/std/the-standard}.
\bibitem{ref_name_cmake}
    CMake. The Cross Platform Build System. Kitware Inc. (2018),
    \url{https://cmake.org}.
\bibitem{ref_name_sympy}
    Meurer, A., Smith, C.~P., Paprocki, M., \v{C}ert\'{i}k, O., Kirpichev, S.~B.,
    Rocklin, M., Kumar, A., Ivanov, S., Moore, J.~K., Singh, S., Rathnayake, T.,
    Vig, S., Granger, B.~E., Muller, R.~P., Bonazzi, F., Gupta, H., Vats, S.,
    Johansson, F., Pedregosa, F., Curry, M.~J., Terrel, A.~R., Rou\v{c}ka, \v{S}.,
    Saboo, A., Fernando, I., Kulal, S., Cimrman, R., Scopatz, A.: SymPy: Symbolic
    Computing in Python. PeerJ Computer Science. \textbf{3}, e103 (2017), doi: 10.7717/peerj-cs.103.
\bibitem{ref_name_python}
    Python Core Team (2015). Python: A Dynamic, Open Source Programming Language.
    Python Software Foundation, \url{https://www.python.org/}.
\bibitem{ref_name_conda} % Vers. 5.1.0.
    Anaconda Software Distribution. Computer Software. Anaconda (2018), \url{https://anaconda.com}.
\bibitem{ref_name_matplotlib}
    Hunter, J.~D.: Matplotlib: A 2D Graphics Environment. Computing in Science
    \& Engineering. \textbf{9}, 3, 90--95 (2007), doi: 10.1109/MCSE.2007.55.
\end{thebibliography}
\end{document}